%% file: 0-main.tex
\documentclass[sigconf]{acmart}
\AtBeginDocument{%
  \providecommand\BibTeX{{%
    \normalfont B\kern-0.5em{\scshape i\kern-0.25em b}\kern-0.8em\TeX}}}

\setcopyright{acmcopyright}
\copyrightyear{2023}
\acmYear{2023}
\acmDOI{XXXXXXX.XXXXXXX}

\acmConference[CHI '23]{ACM CHI Conference on Human Factors in Computing Systems}{April 23--28,
  2023}{Hamburg, Germany}
\acmBooktitle{CHI '23: ACM CHI Conference on Human Factors in Computing Systems,
 April 23--28, 2023, Hamburg, Germany} 
\acmPrice{15.00}
\acmISBN{978-1-4503-XXXX-X/18/06}

\copyrightyear{2023}
\acmYear{2023}
\setcopyright{rightsretained}
\acmConference[CHI EA '23]{Extended Abstracts of the 2023 CHI Conference on Human Factors in Computing Systems}{April 23--28, 2023}{Hamburg, Germany}
\acmBooktitle{Extended Abstracts of the 2023 CHI Conference on Human Factors in Computing Systems (CHI EA '23), April 23--28, 2023, Hamburg, Germany}
\acmDOI{10.1145/3544549.3585903}
\acmISBN{978-1-4503-9422-2/23/04}

\begin{document}

\title[EarCough: Enabling Continuous Subject Cough Event Detection on Hearables]
{EarCough: Enabling Continuous Subject Cough Event Detection on Hearables}

\author{Xiyuxing Zhang}
\email{zxyx22@mails.tsinghua.edu.cn}
\orcid{0009-0002-9337-2278}
\affiliation{%
  \institution{Department of Computer Science and Technology, Tsinghua 
  University}
  \country{China}
}

\author{Yuntao Wang}
\email{yuntaowang@tsinghua.edu.cn}
\authornote{This is the corresponding author.}
\orcid{0000-0002-4249-8893}
\affiliation{%
  \institution{Department of Computer Science and Technology, Tsinghua University}
  \country{China}
}

\author{Jingru Zhang}
\email{zhang-jr19@mails.tsinghua.edu.cn}
\orcid{0009-0000-4336-0151}
\affiliation{%
  \institution{Department of Computer Science and Technology, Tsinghua University}
  \country{China}
}

\author{Yaqing Yang}
\email{yang-yq19@mails.tsinghua.edu.cn}
\orcid{0009-0004-5858-4632}
\affiliation{%
  \institution{Department of Computer Science and Technology, Tsinghua University}
  \country{China}
}

\author{Shwetak Patel}
\email{shwetak@cs.washington.edu}
\orcid{0000-0002-6300-4389}
\affiliation{%
  \institution{University of Washington}
  \country{United States}
}

\author{Yuanchun Shi}
\email{shiyc@tsinghua.edu.cn}
\orcid{0000-0003-2273-6927}
\affiliation{%
  \institution{Department of Computer Science and Technology, Tsinghua University}
  \country{China}
}

\renewcommand{\shortauthors}{Trovato and Tobin, et al.}

\begin{abstract}
Cough monitoring can enable new individual pulmonary health applications. Subject cough event detection is the foundation for continuous cough monitoring. Recently, the rapid growth in smart hearables has opened new opportunities for such needs. This paper proposes EarCough, which enables continuous subject cough event detection on edge computing hearables, by leveraging the always-on active noise cancellation (ANC) microphones. Specifically, we proposed a lightweight end-to-end neural network model --- EarCoughNet. To evaluate the effectiveness of our method, we constructed a synchronous motion and audio dataset through a user study. Results show that EarCough achieved an accuracy of 95.4\% and an F1-score of 92.9\% with a space requirement of only 385 kB. We envision EarCough as a low-cost add-on for future hearables to enable continuous subject cough event detection.
\end{abstract}






\maketitle

\input{1-intro.tex}
\input{2-related.tex}
\input{3-method.tex}
\input{4-res_dis_lim.tex}

\input{5-con_fut.tex}

\begin{acks}
This work is supported by the Natural Science Foundation of China (NSFC) under Grant No.62132010 and No.62002198, Young Elite Scientists Sponsorship Program by CAST under Grant No.2021QNRC001, Tsinghua University Initiative Scientific Research Program, Beijing Key Lab of Networked Multimedia, Institute for Artificial Intelligence, Tsinghua University, and Beijing National Research Center for Information Science and Technology (BNRist).
\end{acks}

\bibliographystyle{ACM-Reference-Format}
\bibliography{refs}

\appendix

\end{document}

%% file: 1-intro.tex
\section{Introduction}
\label{sec:intro}

Cough is one of the most relevant and common indicators of pulmonary diseases\cite{barata2019towards, nemati2020}. Subject cough refers to 
the cough originating from the user of the cough event detection method rather than the cough emitted from the user's environment (such as the cough from other pulmonary disease patients), which is defined as environmental cough. Future automatic cough event detection methods are expected to have subject-awareness~\cite{Hall_2020}, which is the ability to distinguish subject cough events from environmental cough events. Without subject-awareness, cough event detection methods would incorrectly detect environmental coughs and further provide misleading health or disease reports to clinicians, which significantly limits the health application scenarios~\cite{Vatanparvar_2020}. 

In recent years, earphones have become one of the most ubiquitous end-user accessories~\cite{web:canalys}. The global hearables market is projected to reach \$93.90 billion by 2026~\cite{web:alliedmarketresearch}. With the rapid growth of the hearables market, modern smart earphones are developed with rich sensing capabilities and microcontrollers with computational capability, which attracted the research community to explore ways to leverage earphones in the field of health and physiological sensing. Smart earphones have already been leveraged to detect various physiological signals~\cite{Looney2012, Leboeuf2014, Nguyen2016, Bui2019, Athavipach2019, Martin2018, Vogel2007, Roddiger2019a}, including heart rate~\cite{Leboeuf2014, Goverdovsky2016}, brain signals~\cite{Looney2012} and respiration rates~\cite{Roddiger2019a, Liu2019}. Besides, it has been proved that real-time, privacy-safe, low-cost, and ubiquitous cough event detection was able to achieve by leveraging active noise cancellation earphones~\cite{WANG2022}. 

This paper presents EarCough, a technique that enables continuous subject cough event detection on edge computing hearables. Specifically, we proposed a lightweight end-to-end deep learning model named EarCoughNet, which takes the dual-channel audio from hybrid active noise cancellation microphones on smart earbuds as input. To evaluate the effectiveness and reliability of EarCough, we built a dataset targeted at subject cough event detection with sensor fusion data: dual-channel audio data from active noise cancellation microphones of Bose QC 20, plus motion data from the IMU sensor. EarCough achieved an accuracy of 95.4\% and an F1-score of 92.9\% on the constructed dataset, with a space requirement of only 385kB. We envision EarCough as a low-cost add-on for future hearables to enable continuous personal pulmonary health monitoring.

This paper's main contributions have three folds as below.

1) We proposed EarCough, a technique that enables continuous subject cough event detection. To the best of our knowledge, EarCough is the first subject cough event detection method utilizing the difference between dual-channel audio of the built-in always-on hybrid ANC microphones in commodity hearables.

2) We evaluated EarCough's effectiveness and reliability via user study. Results show that EarCough realizes subject cough event detection every 0.5 seconds at an accuracy of 95.4\% and an F1-score of 92.9\% with only 385 kB space requirement.

3) We established the first dataset targeted at continuous subject cough detection with the sensor fusion data: dual-channel audio data from two active noise cancellation microphones plus motion data from the IMU sensor.

%% file: 2-related.tex
\section{Background and Related Works}
\label{sec:rel}

Cough is one of the most common and prominent symptoms associated with many respiratory diseases such as COPD, asthma, and tuberculosis~\cite{barata2019towards,nemati2020}. Automatic cough event detection methods can provide valuable features for pulmonary diagnosis and health condition assessment~\cite{Vatanparvar_2020, Windmon_2019, Hall_2020}. 

Recently, the research community has widely explored automatic cough event detection methods, most of which are audio-based ~\cite{gao2019analysis, Sharan_2019, Liaqat_2021, Teyhouee_2019, birring2008leicester, WANG2022, Xu2021}, owing to the valuable characteristic spectral signature contained in cough sounds~\cite{Amoh_2014}. For instance, Wang et al. ~\cite{WANG2022} proposed HearCough, enabling state-of-the-art continuous cough event detection based on the audio signals from commodity hearables. However, most previous work ignored the importance of distinguishing coughs emitted from the subject (subject coughs) and coughs originating from the subject's environment (environmental coughs), which . In public scenarios, the falsely detected environmental coughs would then be mistakenly considered for a health analysis or disease diagnosis by clinicians, which could have serious adverse consequences due to harmful medication use and increased costs for patients~\cite{Vatanparvar_2020}. 

To fill the gap in subject-awareness, researchers have already explored methods targeting distinguishing between coughs emitted by different coughers. For instance, Whitehill et al. ~\cite{Whitehill_2020} proposed Whosecough, a cougher verification model using audio-based multitask learning strategy and achieves high accuracy under in-the-wild data. Jokic et al. ~\cite{Jokic_2022} presented TripletCough, leveraging triplet network architecture and audio signals captured from smartphones to distinguish cough events among different coughers. However, these works only focus on differentiating the cougher who emits the cough rather than subject cough event detection from other events under different noisy environments, which does not directly feed the need of the healthcare industry. As a result, some previous works introduced subject-awareness into cough event detection methods, which were regarded as subject cough event detection methods. For example, Rahman et al. ~\cite{Rahman2019effi} proposed an audio-based subject cough event detection method, which leveraged feature engineering and random forest, and achieved 94.2\% precision on subject cough event recognition and is suited to be applied on smartphones. Nevertheless, this method was evaluated only on the in-lab dataset, which may suffer from performance drop under in-the-wild scenarios. Besides, no practical approach can be deployed to a microcontroller with only hundreds of kilobytes of RAM (e.g., ARM M4F).

Recently, smart earbuds are often developed more than just audio listening devices, offering an expanding suite of sensors and microcontrollers with computational capability. Most modern earbuds are equipped with active noise cancellation (ANC), which was designed to enhance users' listening experience by reducing environmental noises. Hybrid ANC is one of the most adopted solutions since it produces the best noise reduction while alleviating acoustic discomfort to human ears\footnote{https://blog.teufelaudio.com/hybrid-anc/}. To achieve hybrid active noise cancellation, one or a set of feed-forward (reference) microphones are placed at the outer side of the earphone to collect the environmental noises. Then an adaptive filter running on the digital signal processor will generate an anti-phase signal with $180^{\circ}$ phase delay to the speaker to eliminate the noises propagated to the human ear. A feedback (error) microphone, placed between the speaker and the ear, is deployed to further monitor the noise cancellation performance and then fine-tune the adaptive filter. 

Audio signals received by the feed-forward and feedback microphones are different due to many factors, such as the distance and the orientation of the sound source. As a result, we envision leveraging the difference between the dual-channel audio signals to achieve subject-awareness. Moreover, due to the proximity of earphones to the mouth, high-quality subject cough sounds could be acquired and thus can be leveraged for a more robust cough event detection system. Besides, active noise cancellation earphones are usually compact, portable, and minimally intrusive. As a result, we envision hybrid active noise cancellation smart earbuds as an ideal hardware platform for continuous subject cough event detection.

Compared to previous works, EarCough achieves state-of-the-art subject cough event detection performance with less computational needs. To the best of our knowledge, we are the first to develop a resource-efficient end-to-end subject cough event detection method, which is compatible with the microcontroller in modern earphones.

%% file: 3-method.tex
\section{Method}

\subsection{The Construction of Synchronous Audio and Motion Dataset}
Although in this paper, EarCough takes only audio signals as input, we envision sensor fusion with audio and motion data as a potential method to further improve the performance of subject cough event detection. As a result, we collect synchronous audio and motion data and construct the dataset. 

\subsubsection{Participants}
We first recruited 10 participants (3 females, 7 males) with an average age of 21.4 (s.d. = 0.80). None of them had pulmonary diseases. 

\subsubsection{Hardware platform for synchronous data collection}
\begin{figure}
    \centering
    \includegraphics[width=1\columnwidth]{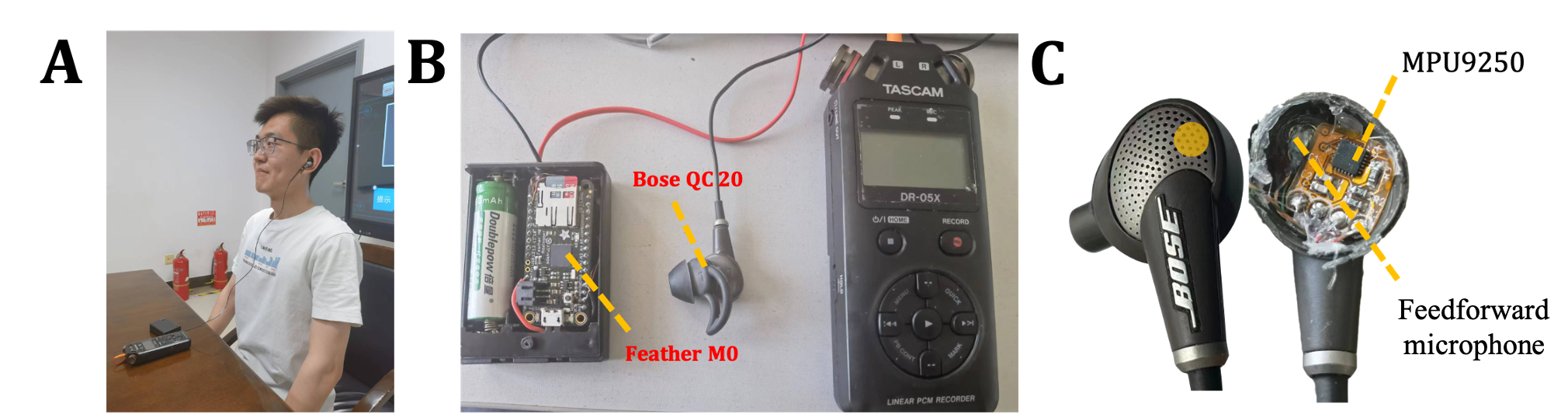}
    \caption{Hardware platform for data collection A: Participant wearing the collection devices B: Three components of the hardware platform C: internal structure of modified Bose QC 20, we embedded the IMU sensor named MPU9250 into the earbuds.}
    \label{fig:coldevice}
    \Description{This figure consists of three subfigures A, B, and C. In subfigure A, a young male participant is seated and wearing data acquisition equipment with earbuds on the left ear. Subfigure B shows the data acquisition equipment, which records the Bose QC 20 audio using a recording pen and collects IMU data through an MPU9250 data cable connected to an SD card-equipped Arduino Feature M0. Subfigure C displays the modified internal structure of the Bose QC 20 earbuds, including the location identification of the feedforward microphone and MPU9250.}
\end{figure}

We established a hardware platform based on smart earbuds for data collection. The hardware platform is shown in Figure ~\ref{fig:coldevice}, which comprises three main components. We used the hybrid ANC earbuds Bose QC 20 as the core collection device, equipped with feed-forward and feedback microphones. Since there is no motion sensor in Bose QC 20, we embedded an inertial measurement unit named MPU9250 into the earbuds. The internal structure of the earbuds is shown in Figure ~\ref{fig:coldevice}C. During the collection, dual-channel audio (48 kHz) is captured by microphones in Bose QC 20 and imported into the DR-05X recorder on the far right of Figure ~\ref{fig:coldevice}B for storage. The embedded MPU9250 shown in Figure ~\ref{fig:coldevice}C is used to collect 6-axis motion signals (including 3-axis accelerometer and 3-axis gyroscope data) at a sampling rate of 1kHz, which are further imported into the Arduino Feather M0 chip on the far left of Figure ~\ref{fig:coldevice}B for processing and storage.

\subsubsection{User study design and procedure}
The user study was conducted in a standard conference room. During the collection procedure, each subject needs to complete the same ten groups of experiments under three sound environments, including quiet room (43 $\sim$ 50dB), noisy room (64 $\sim$70dB), and environmental cough (45 $\sim$ 60dB). The study was approved by the Institutional Review Board (IRB). 

For the noisy indoor sound environment, Bluetooth speakers are used to randomly play all kinds of noise to create the sound environment, which aims to simulate real-life application scenarios. The played background sounds includes natural sound, musical instrument sound, animal sound, human sound, transportation sound and most of the non-cough sound events in users' life. The environmental cough sound environment is created by playing multiple cough audio randomly. All cough audios are high-quality cough audio from FreeSound ~\cite {font2013freesound} website. The sound environment of environmental cough was created to simulate the public social scenarios where multiple people have cough symptoms. We did not simulate two participants coughing simultaneously in the same conference room, which may expose our participants to extremely high health risks.

In each sound environment, the participant wore the hardware and finished seven groups of collections in sitting position, including single cough 10 times, continuous cough 10 times, 5 bites of apple, 5 sips of water, laughing when watching funny videos (video lasts for 1.5 minutes), reading stories for 2 minutes, and randomly move their heads for 1 minute. The participants also finished three groups of collection in walking state, including walking for 30 seconds, single cough 10 times while walking and continuous cough 10 times while walking.

The collecting procedure lasted for about one hour. After completing the data collection user study, a total of 33059.79 seconds of dual-channel audio and motion data were obtained. Each participant received a 15 USD gift card for their time and effort. We passed the hardware prototype among users after thorough sterilization with 75\% alcohol.

To ensure the annotation quality, we recruited three professional data annotators to annotate the data. The statistic results of collected data after annotation was shown in Table ~\ref{tab:dataStat}. 

\begin{table}
\begin{center}
\caption{Statistic results of collected data}
\small
\label{tab:dataStat}
\begin{tabular}{cccc}
 \hline
 \hline
Event   & Total Dur. (sec.) & Average Dur. (sec.) & S.D. \\
 \hline
Single coughs (sitting) & 120.1   & 0.384                        & 0.291                       \\
 \hline
Continuous coughs (sitting) & 247.7                       & 0.796                        & 0.228                       \\ 
 \hline
Bites of apple  & 1519.0                      & 10.264                       & 3.783                       \\ 
 \hline
Sips of water  & 96.2                        & 0.601                        & 0.594                       \\ 
 \hline
Laughing     & 182.3                       & 1.823                        & 1.590                       \\ 
 \hline
Reading     & 2640.7                      & 88.023                       & 10.510                      \\ 
 \hline
Randomly head movement & 631.5                       & 21.050                       & 6.762                       \\ 
 \hline
Walking     & 1036.8                      & 34.560                       & 4.833                       \\ 
 \hline
Single coughs (walking) & 150.3                       & 0.515                        & 0.250                       \\ 
 \hline
Continuous coughs (walking) & 204.0                       & 0.682                        & 0.293                       \\ 
 \hline
Environmental coughs  & 1799.3                      & 1.166                        & 0.323                       \\ 
 \hline
 \hline
\end{tabular}
\end{center}
\end{table}

\subsection{EarCoughNet: An End-to-end Deep Learning Model For Subject Cough Detection}

\subsubsection{Architecture of EarCoughNet}
\begin{figure}[H]
    \centering
    \includegraphics[width=1\columnwidth]{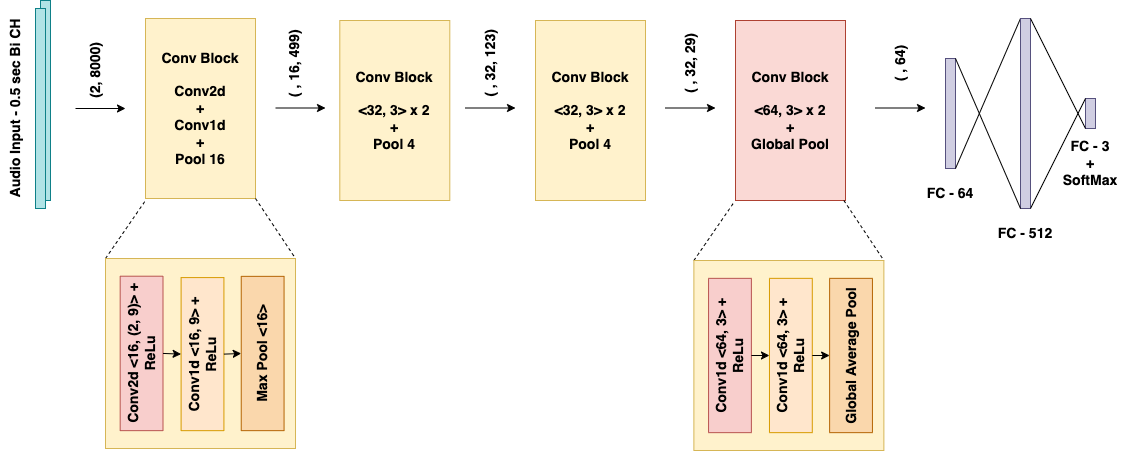}
    \caption{EarCoughNet architecture}
    \label{fig:earcough}
    \Description{This figure shows the network architecture of EarCoughNet, which takes in a 0.5-second dual-channel audio input and passes through four convolution blocks and three fully connected layers. Among them, the first convolution block consists of a two-dimensional convolution layer, a one-dimensional convolution layer, and a max-pooling layer. The second and third convolution blocks both consist of two one-dimensional convolution layers and a max-pooling layer. The final convolution layer consists of two one-dimensional convolution layers and a global pooling layer.}
\end{figure}

The input of EarCoughNet is the raw dual-channel audio signals of the hybrid active noise-canceling earbuds, and the output is the probability of the subject cough event and other events, respectively. As shown in Figure~\ref{fig:earcough}, EarCoughNet consists of four convolution blocks and three fully connected layers. The first convolution layer of the first convolution block is a 2-dimensional convolution layer, which extracts the features of the difference between the dual-channel audio. The rest of the convolutional layers are all 1-dimensional convolutional layers, which has fewer parameters and thus can effectively reduce the model size and resource requirements. 

\subsubsection{Training procedure}
We divided the constructed dataset into training (6 users), validation (2 users) and testing (2 users) dataset. All the audio data was cut into 500-ms audio clips, since cough lasted for 350-ms on average~\cite{amoh2013technologies}. We applied data augmentation methods to the audio of training dataset to expand its size and reduce the model's susceptibility to environmental factors. The audio data augmentation consisted of three stages, which are 

\begin{itemize}
\item Standard audio data augmentation including gain adjustment, time shift, pitch shift, speed adjustment and random masking by making 0-10\% of random points zero. 
\item Noise augmentation including adding machine stimulated white Gaussian noise and mixing background noise from various environmental settings. The background noises are selected from ECS-50~\cite{YDEPUT_2015}.
\item Data formatting including data re-sampling and data normalization.
\end{itemize}

After augmentation, the composition of the three datasets is listed in Table~\ref{tab:division}. We adopted early-stopping strategy on the evaluation dataset to decide the end of our training. We trained four EarCoughNet variants with different sampling rate, including 48kHz, 24kHz, 16kHz, and 8kHz. To evaluate the models' effectiveness, we conducted a cross-user evaluation process on the testing dataset.

\begin{table}[H]
\centering
\caption{Composition of training, validation and testing dataset}
\label{tab:division}
\begin{tabular}{cccc}
\hline
\hline
Samples  & Training   & Validation  & Testing  \\
\hline
Subj. coughs  & 2947 & 842  & 844  \\
\hline
Env. coughs & 4835 & 1208  & 1269  \\
\hline
Other events   & 48104 & 9620 & 10403 \\
\hline
\hline
\multicolumn{4}{l}{\small Audio clips are all 500-ms windows.} \\
\end{tabular}
\end{table}

%% file: 4-res_dis_lim.tex
\section{Results and Discussions}
\subsection{Baselines}
We selected two existed cutting-edge cough event detection methods as baselines. 

\begin{enumerate}
    \item \textbf{HearCough~\cite{WANG2022}.} HearCough is the first effective end-to-end continuous cough event detection method that can be deployed on commodity hearables. However, this work did not consider subject-awareness. We re-implemented and evaluated the model on our constructed dataset.
    \item \textbf{EOCD~\cite{Rahman2019effi}.} EOCD is an efficient online cough detection model deployed on smartphones. This method developed specific modules for subject-awareness and achieved outstanding subject cough event detection performance on in-lab dataset. 
\end{enumerate}

\subsection{Evaluation Metrics}
We considered the following evaluation metrics: 1) detection performance, 2) space requirement, 3) time complexity. 
\begin{itemize}
    \item \textbf{Detection performance.} We used overall accuracy and F-1 score in distinguishing subject coughs and other samples as the major metrics. Further, since subject-awareness is the ability to distinguish subject coughs and environmental coughs, we also used overall accuracy and F-1 score in distinguishing subject coughs and environmental coughs as another pair of performance metrics.
    \item \textbf{Space requirement.} Since the edge computing unit has limited on-board storage that stores the model and the temporary inputs and outputs at each layer, we used space requirement as on-chip metrics, which indicate the deployablility of the techniques.
    \item \textbf{Time complexity.} Real-time detection is an essential ability for cough event detection methods. As a result, less inference time is required, which can be represented by the less time complexity of the model. We used the FLOPs of the model as the metric of time complexity.
\end{itemize}

\subsection{Results and Findings}
We present major evaluation results of all the variants of EarCoughNet and the two baseline methods in Table ~\ref{table:evals} and Table ~\ref{table:evals_2}.

\begin{table}[H]
\begin{center}
 \caption{Evaluation results of effectiveness.}
 \small
 \begin{tabular}{c c c c c c c c c} 
 \hline
 \hline
 \textbf{Model} & \textbf{Input Size} & \textbf{Acc.-1$^+$} & \textbf{F1.-1$^+$} & \textbf{Acc.-2$^\#$} & \textbf{F1.-2$^\#$}\\ 
 \hline
 EarCoughNet & 0.5s @ 48 kHz & 93.37\% & 89.80\% & 96.9\% & 96.8\% \\ 
 \hline
 EarCoughNet & 0.5 s @ 24 kHz & 95.01\% & 92.66\% & 96.4\% & 96.2\% \\ 
 \hline
 EarCoughNet & 0.5 s @ 16 kHz & 95.23\% & 92.25\% & 95.5\% & 96.3\% \\ 
 \hline
 EarCoughNet & 0.5 s @ 8 kHz & 95.35\% & 92.89\% & 95.5\% & 94.6\% \\ 
 \hline
 EOCD$^*$ & 0.6 s @ 44.1 kHz & 94.20\% & 93.70\% & -- & -- \\ 
 \hline
 HearCough & 0.5 s @ 11 kHz & 87.88\% & 88.04\% & 51.64\% & 45.70\%\\ 
 \hline
 \hline
 
\multicolumn{6}{l}{\small $^+$ Acc. and F1. for distinguishing subject coughs and other audio samples.} \\

\multicolumn{6}{l}{\small $^\#$ Acc. and F1. for distinguishing subject coughs and environmental coughs.} \\

\multicolumn{6}{l}{\small $^*$ Unreported values in this row were not evaluated in the original work.} \\

\end{tabular}
\label{table:evals}
\end{center}
\end{table}

\begin{table}[H]
\begin{center}
 \caption{Evaluation results of computing and space requirements.}
 \small
 \begin{tabular}{c c c c c c c c c} 
 \hline
 \hline
 \textbf{Model} & \textbf{Input Size} & \textbf{Flops (M)} & \textbf{Space (kB)}\\
 \hline
 EarCoughNet & 0.5s @ 48 kHz &  73.91 & 1665\\ 
 \hline
 EarCoughNet & 0.5 s @ 24 kHz & 36.88 & 897\\ 
 \hline
 EarCoughNet & 0.5 s @ 16 kHz & 24.53 & 641\\ 
 \hline
 EarCoughNet & 0.5 s @ 8 kHz & 12.20 & 385\\ 
 \hline
 EOCD$^*$ & 0.6 s @ 44.1 kHz & -- & 99000\\ 
 \hline
 HearCough & 0.5 s @ 11 kHz & 16.20 & 480\\ 
 \hline
 \hline
\multicolumn{4}{l}{\small $^*$ Unreported values in this row were not evaluated in the original work.} \\

\end{tabular}
\label{table:evals_2}
\end{center}
\end{table}

\textbf{EarCoughNet is effective for subject cough event detection}. As shown in Table~\ref{table:evals} and ~\ref{table:evals_2}, EarCoughNet achieves comparable detection performance when compared to the EOCD model. However, as an end-to-end model, EarCoughNet has significantly less requirement in computing resources and storage space. Compared to HearCough, EarCoughNet yields higher detection performance at all sampling rates. Worth mentioning, EarCoughNet significantly outperforms HearCough in terms of subject-awareness, which reflects the weakness for works not considering subject-awareness. Since HearCough at an 11 kHz sampling rate can be deployed on the popular microcontroller of commodity hearables, EarCoughNet at 8kHz sampling rate potentially enables real-time on-board subject cough event detection with even less space and computing power compared to HearCough.

\textbf{A sampling rate at 8 kHz is sufficient for subject cough event detection using end-to-end deep learning model}. Coughing sounds have a spectral distribution between 350 Hz and 4 kHz~\cite{Shin_2009}. In order to retent the information of coughs, the sampling rate should be higher than 8kHz according to Nyquist's law. A higher sampling rate provides more high-frequency features, which are unnecessary for cough detection and even introduce more confusing information about cough-like sounds emitted by human beings, including laughter and speaking. According to the analysis of the results, EarCoughNet with a higher sampling rate increases the error rate of misidentifying cough as laughter and speaking, which causes a significant performance drop at the sampling rate of 48kHz. Compared to EarCoughNet with a sampling rate of 16 kHz or 24 kHz, EarCoughNet at the sampling rate of 8 kHz achieves equivalent detection performance with less computing power and space requirement, which showed that 8 kHz is sufficient for subject cough event detection.

\textbf{Effective subject-awareness of EarCoughNet comes from the difference of dual-channel audio}. An ablation study was conducted to investigate the reason for EarCoughNet's outstanding performance in subject-awareness. Specifically, we varied the input of EarCoughNet from dual-channel to single-channel audio from either the feedback microphone or the feed-forward microphone. As shown in ~\ref{tab:fuse_coughres}, single-channel audio-based EarCoughNet suffers from a performance drop in subject-awareness compared to the dual-channel audio-based EarCoughNet, which demonstrated that the difference between the dual-channel audio benefits the performance of subject-awareness.

\begin{table}[H]
\centering
\caption{Ablation study results of EarCoughNet}
\label{tab:fuse_coughres}
\begin{tabular}{ccc}
\hline
\hline
\textbf{Input of EarCoughNet} & \textbf{Acc.$^\#$} & \textbf{F1.$^\#$}\\ 
\hline
Dual-channel audio & 95.50\% & 94.60\% \\ 
\hline
Feed-forward microphone's audio & 88.87\% & 82.15\%  \\ 
\hline
Feedback microphone's audio & 89.77\% & 89.14\% \\ 
\hline
\hline
\multicolumn{3}{l}{\small $^\#$ Acc. and F1. for distinguishing subject coughs and environmental coughs.} \\
\end{tabular}
\end{table}

%% file: 5-con_fut.tex
\section{Conclusion and Future Work}
In this paper, we have proposed EarCough, a technique that enables the always-on hybrid active noise cancellation microphones in commodity hearables for continuous subject cough event detection. We proposed a lightweight end-to-end deep learning model --- EarCoughNet, which leverages the difference between feed-forward and feedback audio of hybrid ANC smart earbuds to achieve effective subject-awareness. To evaluate the effectiveness of the method, we constructed the first synchronous audio and motion dataset targeted at subject cough event detection. We proved EarCough's effectiveness and reliability by training and evaluating the model on the constructed dataset. Results show that EarCough achieved subject cough event detection with an accuracy of 95.4\% and an F1-score of 92.89\% at the audio sampling rate of 8kHz with only 385kB space requirement. To the best of our knowledge, we are the first to develop a resource-efficient end-to-end subject cough event detection compatible with the microcontroller in modern earphones.

Although EarCough achieves outstanding feasibility and efficiency, our constructed dataset needs to expand the evaluation scenarios. For instance, we have not included scenarios when people are listening to the sounds playing from the earbuds. Since the played sounds may influence the quality of the feedback microphone's audio, extra modules based on noise cancellation will be added on EarCough to decrease the negative effect. After expanding the dataset, we will make it publicly accessible to help the research community improve subject cough event detection methods. Besides, we have not evaluated EarCough in real-life scenarios. As a result, we plan to conduct a field study in the wards of stroke patients with cough symptoms and further fine-tune EarCoughNet based on the evaluation results. Finally, since cough usually causes body movements, we envision that sensor fusion methods will further increase the detection performance of EarCough with the motion features extracted from inertial measurement unit (IMU) signals. 